## Title Page

### Full Title

# Lung segmentation on chest x-ray images in patients with severe abnormal findings using deep learning

### Authors and Affiliations

Mizuho Nishio, M.D., Ph.D.[1], Koji Fujimoto, M.D., Ph.D.[2], Kaori Togashi, M.D., Ph.D.[1]

[1] Department of Diagnostic Imaging and Nuclear Medicine, Kyoto University Graduate School of Medicine, 54 Kawahara-cho, Shogoin, Sakyo-ku, Kyoto, Kyoto 606-8507, Japan

[2] Human Brain Research Center, Kyoto University Graduate School of Medicine, 54 Kawahara-cho, Shogoin, Sakyo-ku, Kyoto, Kyoto 606-8507, Japan

*Corresponding author: Mizuho Nishio, M.D., Ph.D.

Department of Diagnostic Imaging and Nuclear Medicine

Kyoto University Graduate School of Medicine

Tel: +81-75-751-3760. Fax: +81-75-771-9709.

e-mail: nishiomizuho@gmail.com, nishio.mizuho.3e@kyoto-u.jp

## Abstract

**Rationale and objectives**: Several studies have evaluated the usefulness of deep learning for lung segmentation using chest x-ray (CXR) images with small- or medium-sized abnormal findings. Here, we built a database including both CXR images with severe abnormalities and experts' lung segmentation results, and aimed to evaluate our network's efficacy in lung segmentation from these images.

**Materials and Methods**: For lung segmentation, CXR images from the Japanese Society of Radiological Technology (JSRT, N = 247) and Montgomery databases (N = 138), were included, and 65 additional images depicting severe abnormalities from a public database were evaluated and annotated by a radiologist, thereby adding lung segmentation results to these images. Baseline U-net was used to segment the lungs in images from the three databases. Subsequently, the U-net network architecture was automatically optimized for lung segmentation from CXR images using Bayesian optimization. Dice similarity coefficient (DSC) was calculated to confirm segmentation.

**Results:** Our results demonstrated that using baseline U-net yielded poorer lung segmentation results in our database than those in the JSRT and Montgomery databases, implying that robust segmentation of lungs may be difficult because of severe abnormalities. The DSC values with baseline U-net for the JSRT, Montgomery and our databases were 0.979, 0.941, and 0.889, respectively, and with optimized U-net, 0.976, 0.973, and 0.932, respectively.

**Conclusion:** For robust lung segmentation, the U-net architecture was optimized via Bayesian optimization, and our results demonstrate that the optimized U-net was more robust than baseline U-net in lung segmentation from CXR images with large-sized abnormalities.

## Introduction

A chest x-ray (CXR) examination is the most widely used diagnostic approach in radiology that is used for screening and diagnosing many types of lung diseases, such as pneumonia, tuberculosis, and lung cancer. Despite the widespread use of this modality in practical clinical situations, it remains difficult for clinicians to accurately detect and diagnose such diseases from CXR images. The accurate assessment of CXR images frequently requires specialized training.

In the case of many types of diseases, there are huge demands for the development of computer-aided diagnosis/detection (CADx/CADe) systems that can aid in diagnosing/detecting them [1–6]. For CXR images, several studies have developed and validated the usefulness and efficacy of CADx/CADe systems [7–10].

High-quality and fast automatic segmentation of organs is a fundamental step for a CADx/CADe system [11]. Lung segmentation has been previously studied in CXR images; however, deformities in the bones, heart, aorta, and pulmonary vein/artery make it difficult to automatically segment the lungs from these images. As a result, a large number of studies that were conducted on the accurate and robust segmentation of the lungs from CXR images were performed using different approaches [8,12–14].

Recently, the use of deep learning has been adopted in medical image analysis and has yielded promising results [1,2,3,8,11,15]. For example, Gordienko et al. [8] used a deep-learning-based model (U-net [15]) for lung segmentation from CXR images. Their U-net-based system successfully segmented the lungs using images from a public database (JSRT database [16]). Although their results were promising, the severity levels of the abnormal findings in the CXR images from the JSRT database were relatively mild. To the best of our knowledge, few studies have evaluated the robustness of a deep-learning-based method in segmenting the lungs from CXR images with severe abnormal findings.

In deep learning, there are many hyperparameters that have to be optimized in order to obtain an accurate and robust model. Although many studies have manually optimized the hyperparameters involved in deep-learning models, the use of Bayesian optimization has resulted in a state-of-the-art performance in tuning the hyperparameters of models [17]. In medical image analysis, one study used Bayesian optimization to optimize the hyperparameters of a CADx system [18]. However, to the best of our knowledge, there are few studies that have used Bayesian optimization to obtain an accurate and robust deep-learning model for medical image analysis.

The purpose of the current study was twofold. (i) The first aim was to use deep learning in order to segment the lungs from CXR images with severe abnormal findings. To validate our system, we built a CXR database for lung segmentation, which included CXR images with severe abnormal findings and their segmentation-related results that were generated by a radiologist. (ii) The second aim of the study was to use Bayesian optimization to optimize the hyperparameters involved in our deep-learning model.

## Materials and Methods

Our study used anonymized data extracted from public databases, and therefore, the regulations in our country did not require us to obtain an institutional review board approval.

### Databases

In this study, we used three different databases, the JSRT database [16], Montgomery database [19], and our own database based on the NIH CXR database [20], to obtain CXR images in order to develop our lung segmentation system and to validate its efficacy. The JSRT database contains 247 CXR images, among which 154 have lung nodules and 93 do not. The results of manual segmentation for the JSRT database are available on the Image Sciences Institute website (http://www.isi.uu.nl/Research/Databases/SCR/) [13]. The Montgomery database contains 138 CXR images, including those of 80 normal patients and of 58 patients diagnosed with tuberculosis. The severity levels of the abnormal findings associated with the lungs were relatively mild in these two datasets. Therefore, in order to evaluate the robustness of our segmentation system, we constructed our own database using CXR images that were selected from the NIH CXR database [20]. A board-certified radiologist with 13 years of experience in diagnostic radiology selected 65 CXR images with severe abnormal findings from the NIH CXR database and manually segmented the lungs from these CXR images.

Because the sizes and image quality of the CXR images were different among the three databases, image preprocessing was performed. The sizes of the CXR images were adjusted to the input size required for the baseline deep-learning model (256×256 pixels). After the CXR images were resized, histogram equalization was performed for normalization.

Deep-learning model

U-net was used as the deep-learning model in the current study, and Bayesian optimization was performed on the baseline U-net model. As a deep-learning model, U-net consists of an encoding–decoding architecture [15]. Further, it comprises of two types of pathways: an analysis pathway for context aggregation and a synthesis pathway that utilizes the semantic and spatial information from the deep layers of U-net. The most important characteristic of U-net is the presence of shortcut connections between the analysis pathway and synthesis pathway at equal resolution. These connections of U-net enable the synthesis pathway to utilize the high-resolution features of the analysis pathway.

U-net is publicly available for implementation in lung segmentation from CXR images [21]. In the current study, we used this publicly available model as the baseline model, and we modified it using Bayesian optimization in order to improve its robustness. The Keras

(https://keras.io/) with Tensorflow (https://www.tensorflow.org/) backends were used for this implementation. Figure 1 presents the U-net baseline model. Both the input and output image dimensions used in the U-net model are 256×256 pixels.

### Training

The Dice loss function was used as the optimization target. Optimization algorithm of U-net is described in the following section. Considering the amount of training data, data augmentation was performed in order to prevent U-net from overfitting. The following settings were implemented for data augmentation: ± 10◦ rotation, ± 10% x-axis shift, ± 10% y-axis shift, and scaling from 80 to 120%. The number of epochs in the training was 100.

### Model modification and hyperparameter optimization

The baseline model presented in Figure 1 was modified to improve the robustness of the segmentation system. The modifications that were made in the current study are as follows. (i) The number of max-pooling layers (*N*) could be changed. (ii) The number of feature maps (*F*) in the first convolution layer could be changed. (iii) In the baseline model, the number of features maps is doubled in each resolution. In our modification, we could select the number of times the number of features maps was doubled (*T*). (iv) Following each convolution layer with a 3×3 kernel, the use of batch normalization could be selected (*BN*) [22]. (v) Following each convolution layer with a 3×3 kernel, the use of dropout and its probability could be selected (*D*). *N, F, T, BN,* and *D* were the hyperparameters of the deep-learning model. In the baseline model, *N* is 4, *F* is 32, *T* is 4, *BN* is 0 (batch normalization is not used), and *D* is 0 (dropout is not used). In addition to the hyperparameters of the deep-learning model, the batch size (*B*), learning rate (*R*), and type of optimization algorithm (*OP*) were also selected as hyperparameters. Therefore, in the current study, eight types of hyperparameters were optimized using Bayesian optimization.

The following hyperparameter space was used in the hyperparameter optimization.

- *B*: 4, 6, 8, 10, 12, 14
- *R*: 0.001, 0.002, 0.003, …, 0.009, 0.01
- *OP*: 1 (Adam), 2 (Nadam), 3 (RMSprop)
- *N*: 3, 4, 5, 6
- *T*: 0, 1, …, *N*
- *F*: 4, 8, 12, 16, …, 40
- *D*: 0, 0.001, 0.002, 0.003, … , 0.019, 0.02 (0 implies that dropout was not used)
- *BN*: 0, 1 (0 implies that batch normalization was not used)

Optuna (https://github.com/pfnet/optuna) was used for the implementation of Bayesian optimization. Optuna searched the hyperparameter space for optimal combinations of the hyperparameters. The optimization target of Optuna was the Dice loss of the test data that was obtained using the modified U-net model. Bayesian optimization was performed using the combined data from the three databases (the total number of CXR images was 450). For Bayesian optimization, 80%, 10%, and 10% of the combined data were used as training data, validation data, and test data, respectively. Since the purpose of the current study was to assess the robustness of our system in segmenting the lungs on CXR images with severe abnormal findings, all the CXR images in the test data were selected from our own database. The number of Bayesian optimization trials was 100.

Evaluation of lung segmentation

After determining the optimal hyperparameters, a quantitative evaluation of lung segmentation was performed for each test dataset from the JSRT and Montogomery databases and our own database. For the quantitative evaluation, a training of the deep-learning model and an evaluation of the model's performance were conducted using data from each database. When the dataset from each of these databases were individually run through the models, the dataset was divided into training data, validation data, and test data. The ratio of splitting that was used was same as what was used for the combined dataset.

The following 4 types of metrics were used for the quantitative evaluation of the segmentation results associated with each database: Dice similarity coefficient (DSC), Jaccard index (JI), sensitivity (SE), and specificity (SP). When calculating these metrics, a threshold of 0.5 was used.

## Results

The results of the 100 Bayesian optimization trials are presented in the supplementary material. Figure 2 presents the changes in test loss during Bayesian optimization. Based on the data presented in the supplementary material and Figure 2, the use of batch normalization had a significant effect on Dice loss in the test data. The three best losses and their hyperparameters are outlined in Table 1. The best Dice loss in the test data from the combined database was 0.0733, which was obtained with the following hyperparameters: $N$ = 4, $T$ = 4, $F$ = 40, $D$ = 0.016, $BN$ = 1, $B$ = 4, $OP$ = 2, and $R$ = 0.001. This optimized model was used for our quantitative evaluation. Figure 3 presents the optimized model obtained by Bayesian optimization.

Table 2 presents the results associated with the four types of metrics for the test data from each of the databases. In the JSRT data, there were almost no differences in results of the four types of metrics between the optimized and baseline models. On the other hand, the results of the optimized model were better than those of the baseline model for the Montogomery database and our database; the DSC values were 0.973 and 0.932 for the optimized model and 0.941 and 0.889 for the baseline model.

Figure 4 presents a representative case with severe abnormal findings. Although a severe abnormal shadow can be identified in Figure 4, the optimized model could robustly segment the lungs compared to the baseline model.

## Discussion

We constructed a CXR database comprising of images with severe abnormal findings to develop a lung segmentation system. Our results reveal that modifying the baseline U-net model via Bayesian optimization was useful in generating a robust and accurate model for lung segmentation. In addition, the lung segmentation results that we acquired via our optimized model from the CXR images with severe abnormal findings were better than those obtained via the baseline model.

Our results demonstrate that the baseline U-net model could be improved via Bayesian optimization. While the network architecture of the baseline U-net model was modified via Bayesian optimization, the modifications were minor in the current study. However, we obtained sufficient improvements in the lung segmentation results from CXR images with severe abnormal findings. These results imply that task-specific hyperparameter tuning was useful in improving the performance of the deep-learning model. Recently, many types of new network architectures have been suggested. To fully utilize the new models, hyperparameter tuning is frequently necessary. Bayesian optimization may be useful for task-specific hyperparameter tuning.

Batch normalization is useful for improving the performance of deep-learning models [22]. Our results obtained using Bayesian optimization demonstrate that the use of batch normalization had a significant effect on segmentation accuracy. While we sometimes encountered the slow convergence of training the U-net model with our segmentation system, the slow convergence could be avoided in the current study by the use of batch normalization. In several cases, batch normalization was useful in dramatically improving the model. Further analysis should be performed to validate the usefulness of batch normalization in developing segmentation systems in medical image analysis.

Based on the data presented in Figure 2, Bayesian optimization selected the use of batch normalization to improve the U-net model; further, the frequency of selecting batch normalization increased in the second half of the Bayesian optimization process. This implies that Bayesian optimization could adaptively select optimal hyperparameters.

While the hyperparameters of the deep-learning model and its training were optimized, several hyperparameters were not optimized in the current study. For example, factors, such as the size of the input image or the use of image preprocessing, were not examined in the current study. Because changes in image size can lead to increased computational costs, we did not evaluate the effect of image size in the current study. However, it may be possible to segment the lungs more precisely in large-sized images. Hyperparameters associated with input should be evaluated in further studies.

There were several limitations to the present study. First, the number of images in our own database was not large. Therefore, our results should be validated using a large-sized database. However, it is difficult to perform manual segmentation to construct a large-sized database. Second, we evaluated the usefulness of Bayesian optimization in optimizing one type of deep-learning model (U-net). Bayesian optimization can be applied to optimize other types of tasks (e.g., classification) and other types of deep-learning models. Future studies should be performed to validate the efficacy of Bayesian optimization in optimizing other tasks or models.

## Conclusion

We constructed a CXR database comprised of images with severe abnormal findings to develop a lung segmentation system for CXR images. Our segmentation system could successfully segment the lungs from CXR images with severe abnormal findings. This robust and accurate model for lung segmentation was obtained by using Bayesian optimization and modifying the U-net model.

**Acknowledgment**

The present study was supported by JSPS KAKENHI (Grant Number JP19K17232). The funder had no role for the present study.

## Figures

Figure 1.

Illustration of the baseline model.

Abbreviations: CXR, chest x-ray; conv, convolution layer; maxpool, max-pooling layer; up conv, up-sampling and convolution layer.

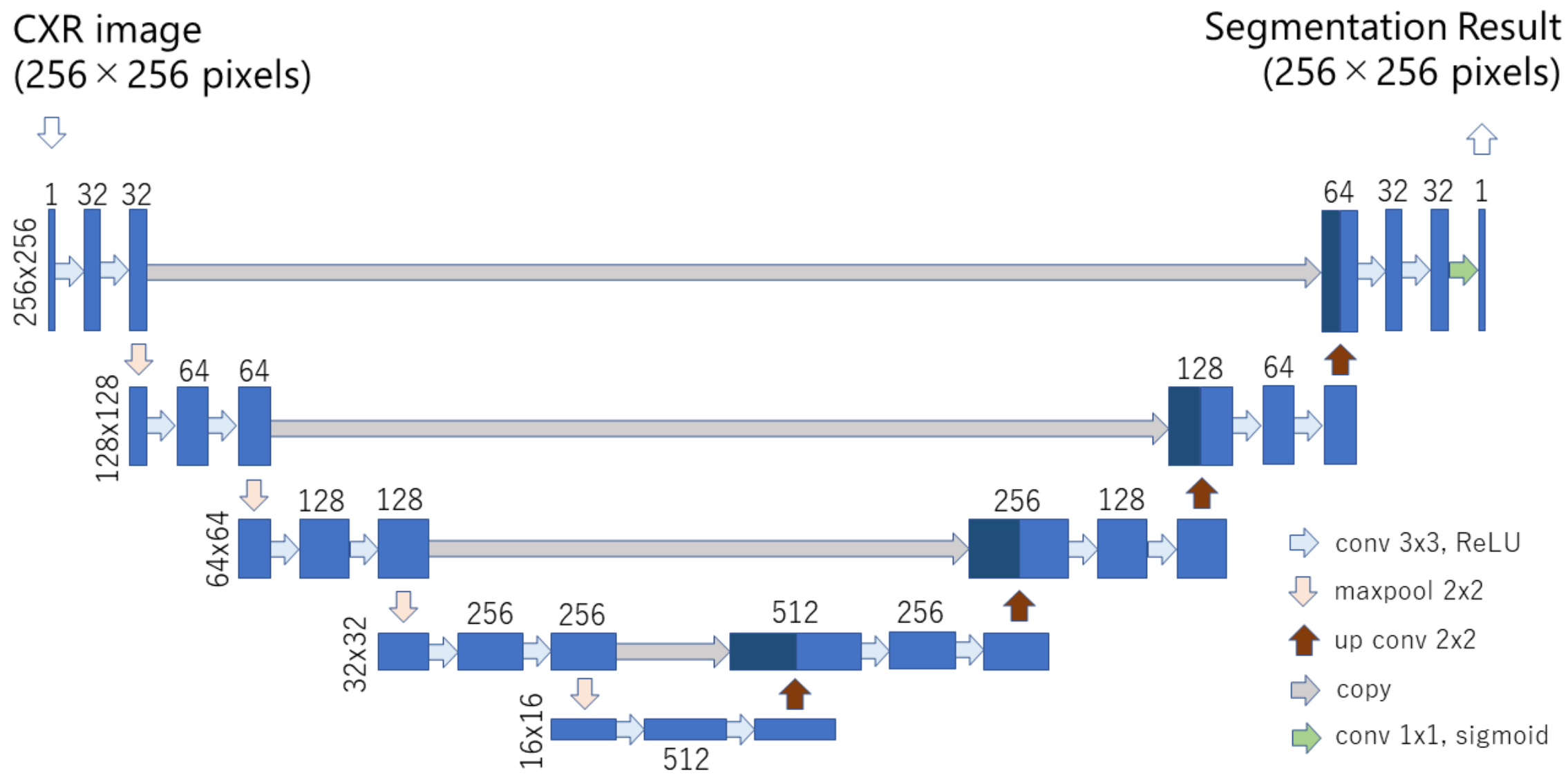

Figure 2.

Change in the Dice loss in the test data during Bayesian optimization.

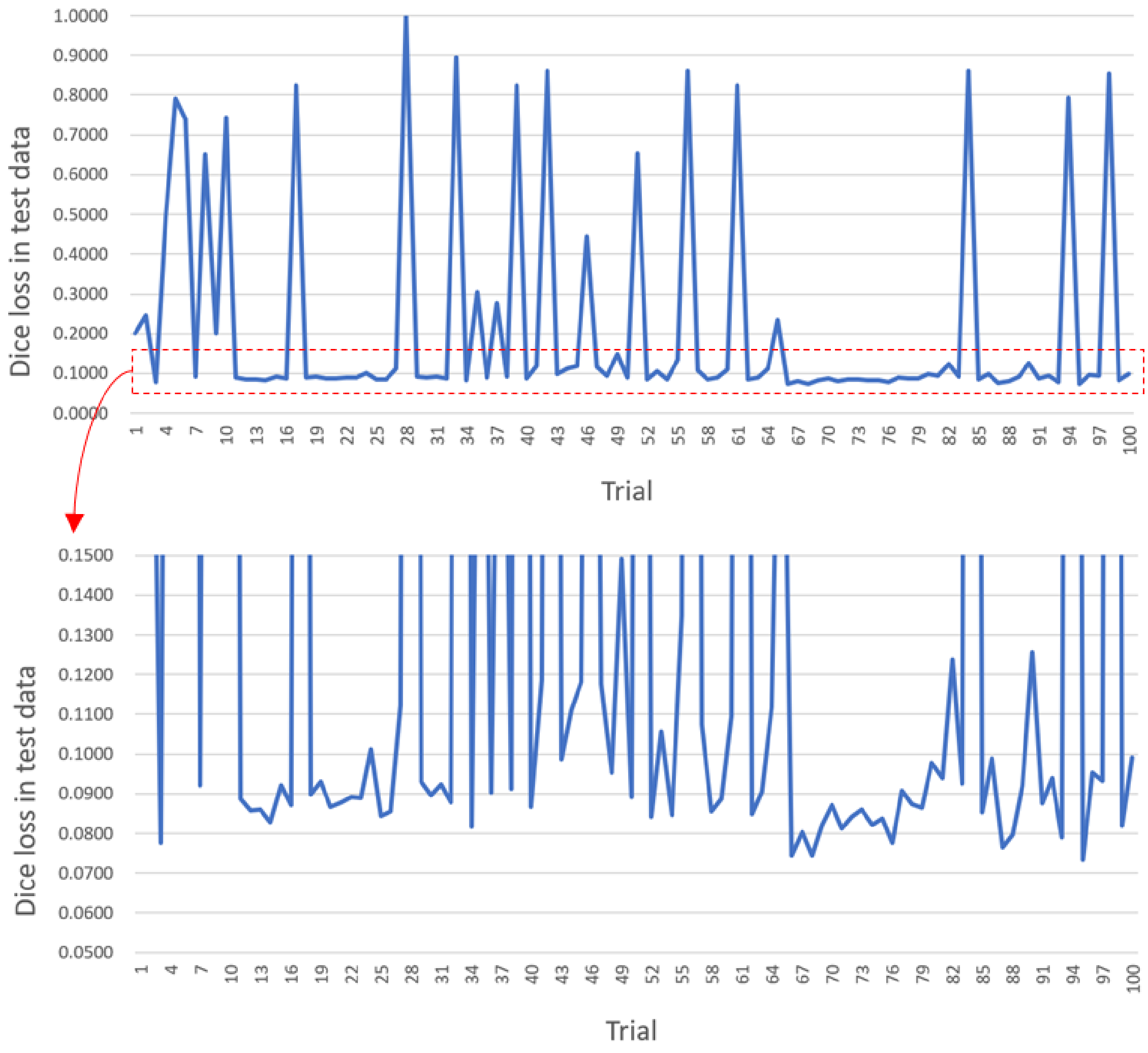

Figure 3.

Illustration of the optimized model.

Abbreviations: CXR, chest x-ray; conv, convolution layer; BN, batch normalization layer; maxpool, max-pooling layer; up conv, up-sampling and convolution layer.

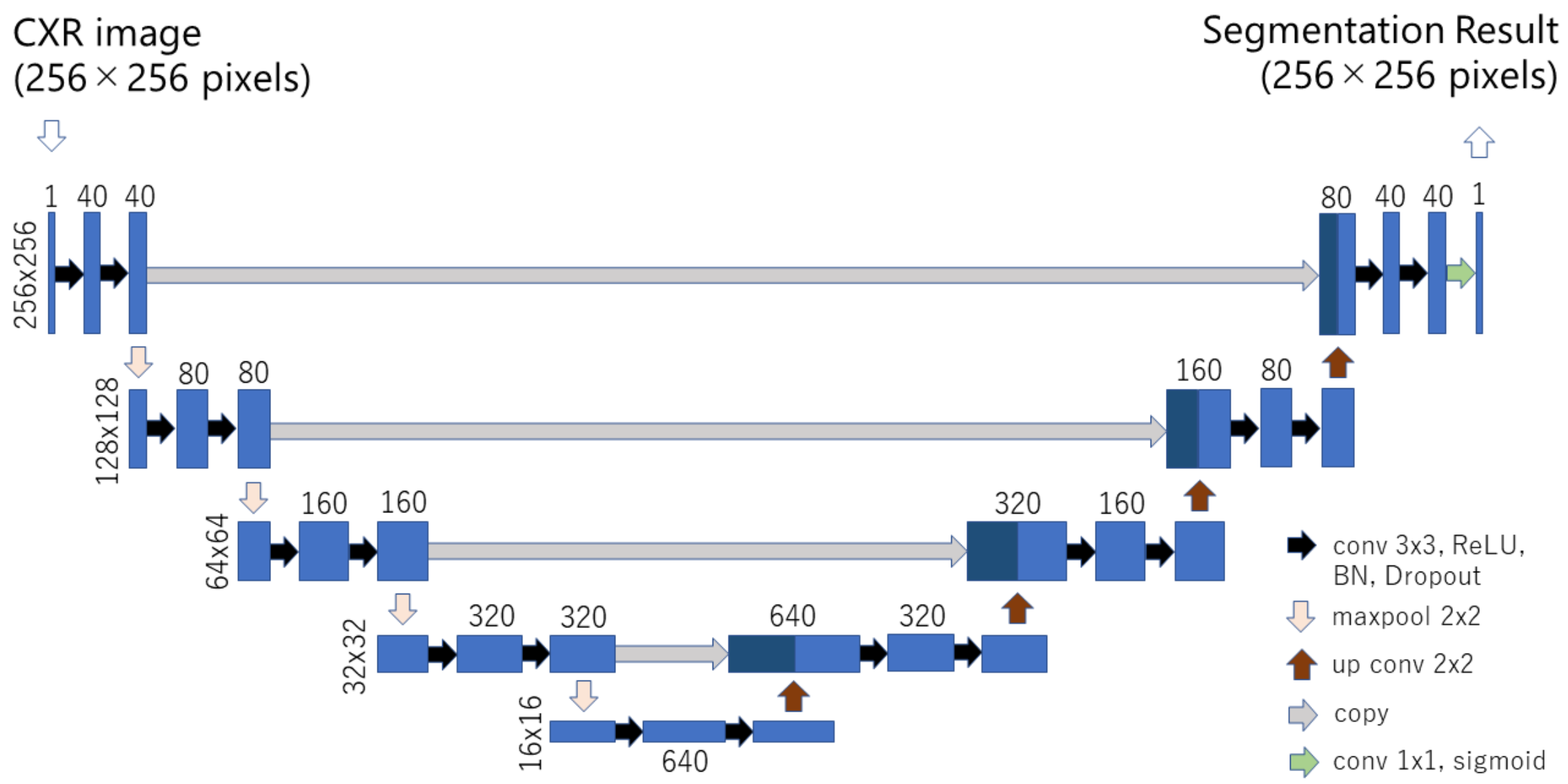

Figure 4.

Chest x-ray image and its segmentation results.

(A)  Chest x-ray image, (B) Ground truth of lung segmentation, (C) Segmentation result of the optimized model, and (D) Segmentation result of the baseline model.

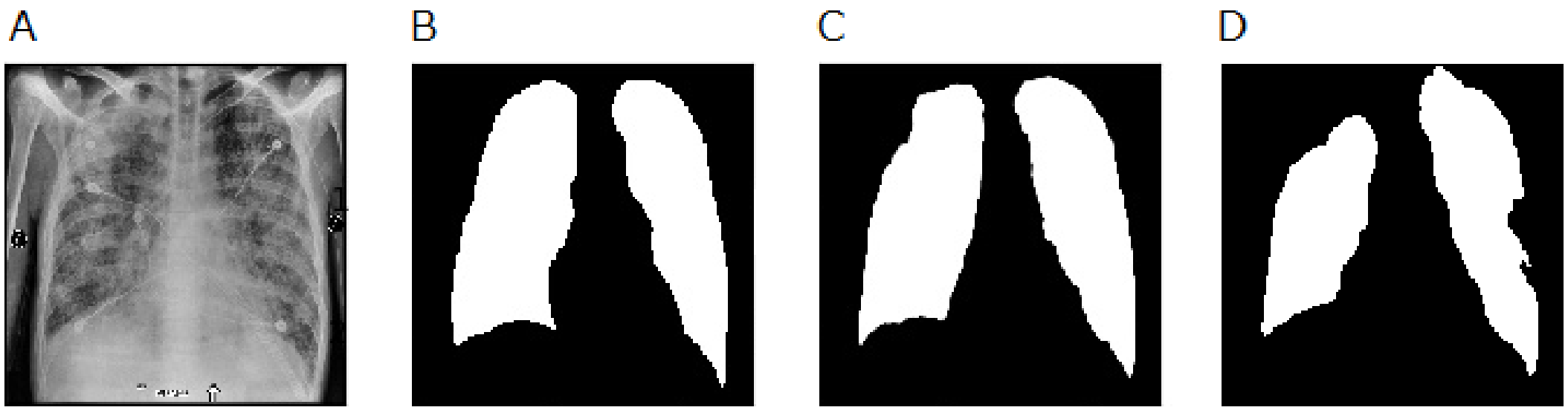

**Tables**

Table 1

Best three losses and their hyperparameters in Bayesian optimization.

| Rank | Dices loss in test data | Hyperparameters |
|---|---|---|
| 1 | 0.0733 | $B$ = 4, $OP$ = 2, $R$ = 0.001, $F$ = 40, $T$ = 4, $N$ = 4, $D$ = 0.016, $BN$ = 1 |
| 2 | 0.0744 | $B$ = 12, $OP$ = 3, $R$ = 0.005, $F$ = 40, $T$ = 4, $N$ = 4, $D$ = 0, $BN$ = 1 |
| 3 | 0.0744 | $B$ = 12, $OP$ = 3, $R$ = 0.005, $F$ = 40, $T$ = 4, $N$ = 4, $D$ = 0, $BN$ = 1 |

Table 2

Results of quantitative evaluation of lung segmentation

| Model | Database | DSC | JI | SE | SP |
|---|---|---|---|---|---|
| Baseline model | JSRT | 0.979 | 0.958 | 0.981 | 0.990 |
| Baseline model | Montogomery | 0.941 | 0.890 | 0.958 | 0.969 |
| Baseline model | our own database | 0.889 | 0.806 | 0.860 | 0.981 |
| Optimized model | JSRT | 0.976 | 0.954 | 0.987 | 0.985 |
| Optimized model | Montogomery | 0.973 | 0.949 | 0.967 | 0.992 |
| Optimized model | our own database | 0.932 | 0.874 | 0.957 | 0.976 |

Abbreviation: Dice similarity coefficient coefficient (DSC), Jaccard index (JI), sensitivity (SE), and specificity (SP).

## Supplementary material

| Trial | Loss | Hyperparameters | | | | | | | |
|---|---|---|---|---|---|---|---|---|---|
| 1 | 0.2013 | B = 10 | OP = 2 | R = 0.009 | F = 4 | T = 0 | N = 3 | D = 0.014 | BN = 1 |
| 2 | 0.2456 | B = 6 | OP = 2 | R = 0.002 | F = 4 | T = 0 | N = 3 | D = 0.006 | BN = 1 |
| 3 | 0.0776 | B = 8 | OP = 1 | R = 0.005 | F = 36 | T = 1 | N = 6 | D = 0.0 | BN = 1 |
| 4 | 0.4924 | B = 12 | OP = 3 | R = 0.002 | F = 12 | T = 0 | N = 3 | D = 0.005 | BN = 0 |
| 5 | 0.7917 | B = 4 | OP = 2 | R = 0.006 | F = 24 | T = 3 | N = 6 | D = 0.002 | BN = 0 |
| 6 | 0.7383 | B = 10 | OP = 1 | R = 0.004 | F = 4 | T = 2 | N = 4 | D = 0.004 | BN = 0 |
| 7 | 0.0921 | B = 14 | OP = 1 | R = 0.007 | F = 20 | T = 2 | N = 4 | D = 0.009 | BN = 1 |
| 8 | 0.6514 | B = 12 | OP = 1 | R = 0.008 | F = 40 | T = 2 | N = 3 | D = 0.009 | BN = 0 |
| 9 | 0.2010 | B = 6 | OP = 1 | R = 0.006 | F = 16 | T = 4 | N = 6 | D = 0.013 | BN = 1 |
| 10 | 0.7430 | B = 12 | OP = 1 | R = 0.003 | F = 4 | T = 3 | N = 6 | D = 0.006 | BN = 0 |
| 11 | 0.0887 | B = 4 | OP = 3 | R = 0.01 | F = 40 | T = 1 | N = 3 | D = 0.018 | BN = 1 |
| 12 | 0.0859 | B = 4 | OP = 2 | R = 0.005 | F = 36 | T = 1 | N = 3 | D = 0.0 | BN = 1 |
| 13 | 0.0859 | B = 4 | OP = 1 | R = 0.001 | F = 36 | T = 1 | N = 3 | D = 0.02 | BN = 1 |
| 14 | 0.0829 | B = 12 | OP = 3 | R = 0.005 | F = 32 | T = 1 | N = 3 | D = 0.0 | BN = 1 |
| 15 | 0.0920 | B = 4 | OP = 2 | R = 0.005 | F = 32 | T = 3 | N = 3 | D = 0.013 | BN = 1 |
| 16 | 0.0871 | B = 4 | OP = 1 | R = 0.009 | F = 32 | T = 0 | N = 3 | D = 0.002 | BN = 1 |
| 17 | 0.8245 | B = 4 | OP = 2 | R = 0.005 | F = 28 | T = 2 | N = 3 | D = 0.016 | BN = 0 |
| 18 | 0.0899 | B = 4 | OP = 3 | R = 0.005 | F = 32 | T = 1 | N = 3 | D = 0.0 | BN = 1 |
| 19 | 0.0930 | B = 4 | OP = 3 | R = 0.005 | F = 28 | T = 1 | N = 3 | D = 0.0 | BN = 1 |
| 20 | 0.0866 | B = 4 | OP = 3 | R = 0.005 | F = 40 | T = 1 | N = 3 | D = 0.003 | BN = 1 |
| 21 | 0.0877 | B = 4 | OP = 3 | R = 0.003 | F = 28 | T = 0 | N = 3 | D = 0.0 | BN = 1 |
| 22 | 0.0892 | B = 12 | OP = 2 | R = 0.007 | F = 40 | T = 1 | N = 3 | D = 0.007 | BN = 1 |
| 23 | 0.0889 | B = 4 | OP = 3 | R = 0.01 | F = 40 | T = 2 | N = 3 | D = 0.002 | BN = 1 |
| 24 | 0.1012 | B = 4 | OP = 2 | R = 0.008 | F = 28 | T = 1 | N = 3 | D = 0.008 | BN = 1 |
| 25 | 0.0844 | B = 4 | OP = 3 | R = 0.001 | F = 24 | T = 0 | N = 3 | D = 0.011 | BN = 1 |
| 26 | 0.0856 | B = 4 | OP = 1 | R = 0.004 | F = 40 | T = 2 | N = 3 | D = 0.004 | BN = 1 |
| 27 | 0.1123 | B = 4 | OP = 2 | R = 0.005 | F = 40 | T = 1 | N = 3 | D = 0.011 | BN = 1 |
| 28 | 1.0000 | B = 4 | OP = 3 | R = 0.005 | F = 16 | T = 0 | N = 3 | D = 0.001 | BN = 0 |
| 29 | 0.0930 | B = 4 | OP = 2 | R = 0.009 | F = 40 | T = 1 | N = 3 | D = 0.004 | BN = 1 |
| 30 | 0.0896 | B = 4 | OP = 1 | R = 0.005 | F = 40 | T = 0 | N = 3 | D = 0.015 | BN = 1 |
| 31 | 0.0922 | B = 12 | OP = 2 | R = 0.002 | F = 40 | T = 2 | N = 3 | D = 0.006 | BN = 1 |
| 32 | 0.0878 | B = 4 | OP = 3 | R = 0.001 | F = 40 | T = 0 | N = 3 | D = 0.001 | BN = 1 |
| 33 | 0.8955 | B = 4 | OP = 2 | R = 0.006 | F = 12 | T = 1 | N = 3 | D = 0.005 | BN = 0 |
| 34 | 0.0818 | B = 12 | OP = 1 | R = 0.002 | F = 40 | T = 3 | N = 3 | D = 0.003 | BN = 1 |
| 35 | 0.3043 | B = 12 | OP = 1 | R = 0.002 | F = 4 | T = 3 | N = 3 | D = 0.003 | BN = 0 |
| 36 | 0.0904 | B = 4 | OP = 1 | R = 0.002 | F = 40 | T = 3 | N = 3 | D = 0.008 | BN = 1 |
| 37 | 0.2763 | B = 4 | OP = 1 | R = 0.002 | F = 4 | T = 3 | N = 3 | D = 0.005 | BN = 0 |
| 38 | 0.0913 | B = 12 | OP = 1 | R = 0.007 | F = 40 | T = 3 | N = 3 | D = 0.01 | BN = 1 |
| 39 | 0.8246 | B = 4 | OP = 1 | R = 0.004 | F = 40 | T = 2 | N = 3 | D = 0.002 | BN = 0 |
| 40 | 0.0867 | B = 12 | OP = 1 | R = 0.008 | F = 40 | T = 2 | N = 3 | D = 0.007 | BN = 1 |
| 41 | 0.1186 | B = 4 | OP = 1 | R = 0.002 | F = 4 | T = 3 | N = 3 | D = 0.003 | BN = 1 |
| 42 | 0.8613 | B = 12 | OP = 1 | R = 0.003 | F = 40 | T = 2 | N = 3 | D = 0.001 | BN = 0 |
| 43 | 0.0987 | B = 4 | OP = 1 | R = 0.01 | F = 4 | T = 2 | N = 3 | D = 0.005 | BN = 1 |
| 44 | 0.1112 | B = 12 | OP = 1 | R = 0.006 | F = 4 | T = 3 | N = 3 | D = 0.02 | BN = 1 |
| 45 | 0.1182 | B = 4 | OP = 2 | R = 0.002 | F = 4 | T = 2 | N = 3 | D = 0.013 | BN = 1 |
| 46 | 0.4439 | B = 12 | OP = 1 | R = 0.009 | F = 4 | T = 0 | N = 3 | D = 0.017 | BN = 0 |
| 47 | 0.1178 | B = 4 | OP = 2 | R = 0.004 | F = 4 | T = 1 | N = 3 | D = 0.007 | BN = 1 |
| 48 | 0.0952 | B = 12 | OP = 1 | R = 0.007 | F = 4 | T = 3 | N = 3 | D = 0.009 | BN = 1 |
| 49 | 0.1491 | B = 12 | OP = 1 | R = 0.008 | F = 4 | T = 2 | N = 3 | D = 0.001 | BN = 1 |
| 50 | 0.0892 | B = 4 | OP = 2 | R = 0.003 | F = 40 | T = 1 | N = 3 | D = 0.004 | BN = 1 |
| 51 | 0.6528 | B = 12 | OP = 1 | R = 0.01 | F = 40 | T = 2 | N = 3 | D = 0.012 | BN = 0 |
| 52 | 0.0841 | B = 4 | OP = 2 | R = 0.006 | F = 40 | T = 0 | N = 3 | D = 0.019 | BN = 1 |
| 53 | 0.1056 | B = 12 | OP = 1 | R = 0.001 | F = 40 | T = 3 | N = 3 | D = 0.006 | BN = 1 |
| 54 | 0.0846 | B = 4 | OP = 1 | R = 0.002 | F = 40 | T = 1 | N = 3 | D = 0.008 | BN = 1 |
| 55 | 0.1350 | B = 12 | OP = 2 | R = 0.009 | F = 4 | T = 1 | N = 3 | D = 0.003 | BN = 1 |
| 56 | 0.8616 | B = 4 | OP = 1 | R = 0.005 | F = 40 | T = 0 | N = 3 | D = 0.01 | BN = 0 |
| 57 | 0.1076 | B = 12 | OP = 2 | R = 0.007 | F = 40 | T = 2 | N = 3 | D = 0.0 | BN = 1 |
| 58 | 0.0856 | B = 12 | OP = 1 | R = 0.003 | F = 40 | T = 1 | N = 3 | D = 0.002 | BN = 1 |
| 59 | 0.0887 | B = 12 | OP = 2 | R = 0.008 | F = 40 | T = 1 | N = 3 | D = 0.015 | BN = 1 |
| 60 | 0.1096 | B = 12 | OP = 1 | R = 0.004 | F = 4 | T = 3 | N = 3 | D = 0.006 | BN = 1 |
| 61 | 0.8244 | B = 12 | OP = 2 | R = 0.005 | F = 40 | T = 2 | N = 3 | D = 0.011 | BN = 0 |
| 62 | 0.0848 | B = 12 | OP = 1 | R = 0.01 | F = 40 | T = 0 | N = 3 | D = 0.001 | BN = 1 |
| 63 | 0.0906 | B = 12 | OP = 2 | R = 0.001 | F = 40 | T = 1 | N = 3 | D = 0.003 | BN = 1 |
| 64 | 0.1119 | B = 12 | OP = 1 | R = 0.002 | F = 40 | T = 3 | N = 3 | D = 0.004 | BN = 1 |
| 65 | 0.2349 | B = 12 | OP = 1 | R = 0.005 | F = 4 | T = 1 | N = 3 | D = 0.002 | BN = 0 |
| 66 | 0.0744 | B = 12 | OP = 3 | R = 0.005 | F = 40 | T = 4 | N = 4 | D = 0.0 | BN = 1 |
| 67 | 0.0803 | B = 12 | OP = 3 | R = 0.005 | F = 40 | T = 4 | N = 4 | D = 0.0 | BN = 1 |
| 68 | 0.0744 | B = 12 | OP = 3 | R = 0.005 | F = 40 | T = 4 | N = 4 | D = 0.0 | BN = 1 |
| 69 | 0.0822 | B = 12 | OP = 3 | R = 0.005 | F = 40 | T = 3 | N = 3 | D = 0.0 | BN = 1 |
| 70 | 0.0872 | B = 12 | OP = 3 | R = 0.005 | F = 40 | T = 0 | N = 3 | D = 0.001 | BN = 1 |
| 71 | 0.0813 | B = 12 | OP = 3 | R = 0.005 | F = 40 | T = 1 | N = 3 | D = 0.002 | BN = 1 |
| 72 | 0.0841 | B = 12 | OP = 3 | R = 0.005 | F = 40 | T = 3 | N = 3 | D = 0.004 | BN = 1 |
| 73 | 0.0861 | B = 12 | OP = 3 | R = 0.005 | F = 40 | T = 0 | N = 3 | D = 0.005 | BN = 1 |
| 74 | 0.0821 | B = 12 | OP = 3 | R = 0.005 | F = 40 | T = 1 | N = 3 | D = 0.001 | BN = 1 |
| 75 | 0.0837 | B = 12 | OP = 3 | R = 0.005 | F = 40 | T = 4 | N = 4 | D = 0.0 | BN = 1 |
| 76 | 0.0776 | B = 12 | OP = 3 | R = 0.006 | F = 40 | T = 4 | N = 4 | D = 0.003 | BN = 1 |
| 77 | 0.0907 | B = 12 | OP = 3 | R = 0.006 | F = 4 | T = 3 | N = 3 | D = 0.002 | BN = 1 |
| 78 | 0.0873 | B = 4 | OP = 3 | R = 0.006 | F = 40 | T = 3 | N = 3 | D = 0.009 | BN = 1 |
| 79 | 0.0864 | B = 4 | OP = 3 | R = 0.006 | F = 40 | T = 3 | N = 3 | D = 0.005 | BN = 1 |
| 80 | 0.0979 | B = 4 | OP = 3 | R = 0.006 | F = 40 | T = 3 | N = 3 | D = 0.007 | BN = 1 |
| 81 | 0.0940 | B = 14 | OP = 3 | R = 0.009 | F = 4 | T = 3 | N = 3 | D = 0.003 | BN = 1 |
| 82 | 0.1239 | B = 4 | OP = 3 | R = 0.006 | F = 4 | T = 3 | N = 3 | D = 0.001 | BN = 1 |
| 83 | 0.0927 | B = 14 | OP = 3 | R = 0.007 | F = 4 | T = 3 | N = 3 | D = 0.006 | BN = 1 |
| 84 | 0.8609 | B = 4 | OP = 3 | R = 0.004 | F = 40 | T = 3 | N = 3 | D = 0.0 | BN = 0 |
| 85 | 0.0852 | B = 14 | OP = 3 | R = 0.008 | F = 40 | T = 3 | N = 3 | D = 0.004 | BN = 1 |
| 86 | 0.0989 | B = 4 | OP = 3 | R = 0.003 | F = 40 | T = 3 | N = 3 | D = 0.014 | BN = 1 |
| 87 | 0.0765 | B = 12 | OP = 3 | R = 0.001 | F = 40 | T = 3 | N = 3 | D = 0.008 | BN = 1 |
| 88 | 0.0796 | B = 12 | OP = 3 | R = 0.001 | F = 40 | T = 4 | N = 4 | D = 0.017 | BN = 1 |
| 89 | 0.0922 | B = 4 | OP = 3 | R = 0.001 | F = 4 | T = 4 | N = 4 | D = 0.012 | BN = 0 |
| 90 | 0.1257 | B = 14 | OP = 3 | R = 0.001 | F = 4 | T = 4 | N = 4 | D = 0.014 | BN = 1 |
| 91 | 0.0875 | B = 4 | OP = 3 | R = 0.001 | F = 4 | T = 4 | N = 4 | D = 0.008 | BN = 1 |
| 92 | 0.0940 | B = 12 | OP = 2 | R = 0.001 | F = 4 | T = 4 | N = 4 | D = 0.019 | BN = 1 |
| 93 | 0.0790 | B = 4 | OP = 3 | R = 0.01 | F = 40 | T = 4 | N = 4 | D = 0.012 | BN = 1 |
| 94 | 0.7929 | B = 12 | OP = 3 | R = 0.009 | F = 40 | T = 5 | N = 5 | D = 0.009 | BN = 0 |
| 95 | 0.0733 | B = 4 | OP = 2 | R = 0.001 | F = 40 | T = 4 | N = 4 | D = 0.016 | BN = 1 |
| 96 | 0.0953 | B = 4 | OP = 2 | R = 0.007 | F = 40 | T = 3 | N = 3 | D = 0.017 | BN = 1 |
| 97 | 0.0932 | B = 6 | OP = 2 | R = 0.008 | F = 4 | T = 3 | N = 3 | D = 0.016 | BN = 1 |
| 98 | 0.8533 | B = 4 | OP = 2 | R = 0.004 | F = 4 | T = 3 | N = 3 | D = 0.019 | BN = 0 |
| 99 | 0.0819 | B = 4 | OP = 2 | R = 0.003 | F = 40 | T = 2 | N = 3 | D = 0.018 | BN = 1 |
| 100 | 0.0992 | B = 4 | OP = 2 | R = 0.01 | F = 4 | T = 2 | N = 3 | D = 0.02 | BN = 1 |